\documentclass{epl}
\usepackage{graphicx}
\usepackage{enumerate}
\usepackage{amsmath}
\usepackage{amssymb}

\title{Induced order and reentrant melting in classical two-dimensional binary clusters}

\author{K. Nelissen\inst{1} \and B. Partoens\inst{1} \and I. Schweigert\inst{1,2} \and F. M. Peeters \inst{1}}
\institute{
  \inst{1} Departement Fysica,Universiteit Antwerpen, Groenenborgerlaan 171, B-2020 Antwerpen,Belgium\\
  \inst{2} Institute of Theoretical and Applied Mechanics, Institutskaya 4/1, Novosibirsk \\
  630090, Russia }

\begin{document}

\maketitle


\begin{abstract}
A binary system of classical charged particles interacting through
a dipole repulsive potential and confined in a two-dimensional
hardwall trap is studied by Brownian dynamics simulations. We found that the presence of small
particles \emph{stabilizes} the angular order of the system as a
consequence of radial fluctuations of the small particles. There
is an optimum in the increased rigidity of the cluster as function
of the number of small particles. The small (i.e. defect)
particles melt at a lower temperature compared to the big
particles and exhibit a \emph{reentrant} behavior in its radial
order that is induced by the intershell rotation of the big
particles.
\end{abstract}

\section{Introduction}

Melting and crystallization are fundamental processes in nature
and have been widely studied. Charged particles systems like
e.g.~colloids ~\cite{pusey} and dusty plasma's~\cite{chu} display
similar phase behavior as atoms and molecules with the added
advantage that the micrometer size of the particles and their
slower dynamics make them accessible for real space imaging
\cite{murray}. Most of the previous research was directed towards
one-component systems. Recently, in a theoretical study in Ref.
\cite{drocco} the complexity of the system was increased by investigating
systems with two types of particles of different radii and
different effective charge confined in a parabolic trap. A recent
experimental study \cite{leunissen} concentrated on oppositely charged
colloidal particles confined in a cavity and found a remarkable
diversity of new binary structures. In this letter we consider a
finite size binary system of repulsive particles which are
confined to move in two dimensions (2D). The circular hard wall
confinement potential competes with the 2D Wigner crystal
structure~\cite{wigner} and leads to ring like arrangements for
the particles~\cite{Grimes, Bedanov}. Previously it was shown
experimentally~\cite{bubeck} and theoretically~\cite{schweigert}
that single component systems exhibit a remarkable re-entrant
melting behavior. In the present binary system we assume a large
difference in the size and `charge' of the particles and therefore
the smaller particles can be considered as `defects' which disturb
the order of the big particles~\cite{kwinten}. We found that these defect
particles have a pronounced effect on the melting behavior of the
system and results in an unexpected stabilization of the ordered
phase and a new reentrant melting behavior. The possibility of
stabilization was also addressed in the theoretical study of Ref. \cite{drocco} in the case of few binary Coulomb clusters confined in a
parabolic trap.

The present study is motivated by the experiment of
Ref.~\cite{mangold} where the melting behavior of a binary system
of paramagnetic colloidal spheres (with different radii) confined
in 2D circular cavities was studied. The coupling parameter could
be tuned by changing the applied magnetic field strength. They
found that: 1) the shell-like structure of the system depends
strongly on the relative number of big ($N_B$) and small particles
($N_S$), and 2) the melting process takes place in several stages
where first the small particles and afterwards the big particles
become delocalized.

\section{Model system}

In our model, like the experiment of Ref.~\cite{mangold}, the
particles are confined by a circular hard wall potential ($V_{P} =
0$ for $r \leq R$ and $V_{p} = \infty$ at $r > R$). Like in the
experiment, the particles interact through a repulsive dipole
potential
$V(\vec{r}_{i},\vec{r}_{j})=q_i.q_j/|\vec{r}_{i}-\vec{r}_j|^3$,
where $q_i=M_i \sqrt{m_0/4\pi}$ is the `charge' and $\vec{r}_i$
the coordinate of the $i^{\emph{th}}$ particle. For a given type
of interparticle interaction and external confinement, only three
parameters characterize the order of the system: the number of big
particles \emph{$N_B$}, the number of small particles $N_S$ (also called
defect particles) and
the coupling parameter $\Gamma$. In the experiment the diameter of the big particles is twice the diameter of the small particles~\cite{mangold}, therefore we have
chosen the charge of the small particles to be $1/8^{th}$ of the
charge of the big particles. As a representative example we will
discuss in the following the results for clusters with 16 and 17
big particles. We define the characteristic energy of the
inter-particle interaction for dipole clusters as $E_0 = q^2 /
a_0^3$, where $a_0 = 2 R/{N_B}^{1/2}$ is the average distance
between the big particles. In the present calculation we define
the coupling parameters as $\Gamma = q^2 / a_0^3 k_B T$, where
$k_B$ is the Boltzmann constant and $T$ the temperature of the
medium.

The ratio of the particle velocity relaxation time versus the
particle position relaxation time is very small due to the
viscosity of water and consequently the motion of the particles is
diffusive and overdamped. In our simulations we will neglect hydrodynamic
interactions. Following \cite{ermak} we rewrite the stochastic
Langevin equations of motion for the position of the particles as
those for Brownian particles:
\begin{equation}
\frac{d\vec{r_i}}{dt}=\frac{D_i}{k_BT}\left\{ \sum_{j=1}^N
\frac{dV(\vec{r_i},\vec{r_j})}{d\vec{r}} +
\frac{dV_P(\vec{r_i})}{d\vec{r}} + \frac{\vec{F_L^i}}{m_{i}}
\right\},
\end{equation}
where $D_i$ is the self-diffusion coefficient and $m_i$ is the
particle mass of the $i^{th}$ particle, and $\vec{F_L^i}$ is the
randomly fluctuating force acting on the particles due to the
surrounding media. In the numerical solution of Eq. (1) we took a
time step $\Delta t \leq 10^{-4}/(nD_B)$, where $n = N_B/(\pi
R^2)$ is the density of the big particles. The radius of the
circular vessel is $R = 36$ $\mu$m  and the self-diffusion
coefficient of the big particles $D_B = 0.35$ $\mu$m$^2$/s is
taken from the experiment \cite{mangold}. As the self-diffusion
constant is inversely proportional to the particle diameter (from
the Stokes-Einstein relation $D=k_{B}T/8\pi \nu a$, with
\emph{$\nu$} the viscosity of the fluid and \emph{a} the particle
diameter) we took $D_S=0.7\mu m^2 /s$ for the small particles.

Before starting the Brownian dynamics we find first the
groundstate configuration using the Monte Carlo technique as in
Ref.~\cite{schweigert95}.

In order to characterize the angular order of the system, we
calculate the angular diffusion of the particles over a 30 min x
1000 time interval. The relative angular diffusion coefficient can
be written as
\begin{equation}
D_\theta = \left\{\langle\Delta \theta (t)^2\rangle -
\langle\Delta \theta (t)\rangle^2\right\}/t,
\end{equation}
where $\langle\rangle$ refers to a time averaging, and the mean
relative angular displacement rotation of the first shell
$[\theta_1 (t)]$ relative to the second $[\theta_2 (t)]$ one is
defined as $\Delta \theta (t) = [\theta_2 (t)] - [\theta_1 (t)]$.
The mean squared radial diffusion (MSRD) coefficient is
\begin{equation}
\Delta R^2 = \frac{1}{N} \sum_{i=1}^N [\langle r_i(t)^2\rangle -
\langle r_i(t)\rangle^2]/a_0,
\end{equation}
which is a measure of the radial order in the system. The MSRD is
calculated separately for the big and the small particles.

\section{Results}

\begin{figure}
\centering
\resizebox{20pc}{!}{\includegraphics{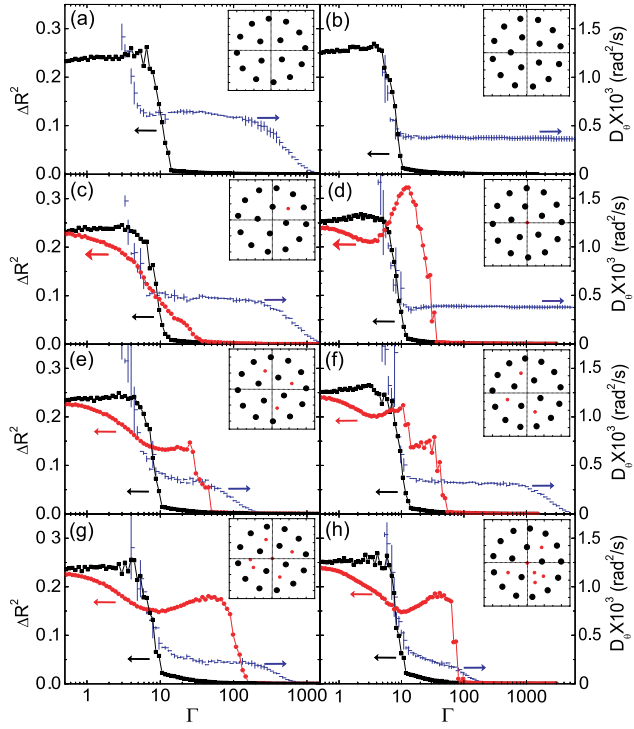}}
\caption{(color online) Left column: system containing $16$
particles. From top to bottom the number of small particles
increases from 0, 1, 3, 6. Right column: system containing $17$
particles. Right scale: the relative angular diffusion coefficient
(given with error bars) as function of $\Gamma$. Left scale: the
$\Delta R^2$ for the small particles (red bullets) and the big
particles (black squares) in the binary cluster as function of
$\Gamma$. The insets show the corresponding groundstate
configuration at $T=0$} \label{res}
\end{figure}

We found that all the interesting melting properties for small
binary clusters are present in the $N_B=16$ and $N_B=17$ systems.
In the insets of Fig.~\ref{res} the ground state configurations
for these systems are shown with zero ((a) and (e)), one ((b) and
(f)), three ((c) and (g)) and six ((d) and (h)) small particles.
In a $16$-particle system with less than $9$ small particles (left
column in Fig.~\ref{res}), one can see that the big particles form
a shell structure with $4$ particles in the inner shell and $12$
particles localized at the edge of the hard wall. The small
particles fill up the vacancies between the big particles. The
difference of charge between the big particles and the small
particles is so large, that the small particles are expelled from
the outer ring. The (4,12)-configuration is a magic number
configuration and is exactly the same configuration as one finds
without small particles~\cite{minghui}. (For $9$ or more small
particles the magic configuration is lost and the big particles
form a non-magic configuration.) In a $17$-particle system (right
column in Fig.~\ref{res}) with no small particles, the big
particles form the (5,12)-configuration. However, by adding 2
small particles to the system, the ground state configuration of
the big particles changes into the (4,13)-configuration. The
reason for this change in the ground state is that the
(5,12)-configuration is a non-magic number configuration and by
adding small particles the system tries to adjust to a more
triangular lattice.

In order to study the melting of the binary clusters we performed
Brownian dynamics simulations for several values of the coupling
constant $\Gamma$. First we show how the angular melting properties change by adding small particles. Afterwards we deal with the radial melting properties. 

\subsection{Angular melting}

\begin{figure}
\centering
\includegraphics[width=250pt]{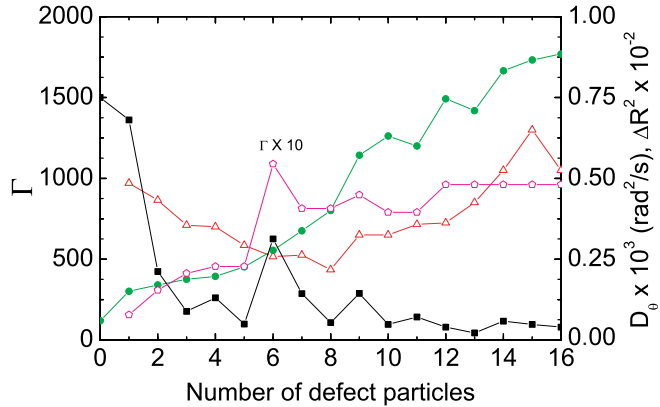} \caption{(color online) System containing $16$ particles. Black squares: critical $\Gamma$
where the angular diffusion of the big particles is exceeding the
value $100$. Red triangles: height of the plateau of the angular
diffusion curve measured at $\Gamma=20$. Green dots: height of
the $\Delta R^2$ of the big particles measured at $\Gamma=20$.
Pink pentagons: Melt temperature of the small particles. }
\label{16}
\end{figure}

The relative angular diffusion coefficients as
function of $\Gamma$ for a system with $16$ and $17$ big particles
for different number of small particles are shown in
Fig.~\ref{res}. We notice from Fig.~\ref{res}(a), for the magic number cluster $(4,12)$ without small particles, that the
relative angular diffusion curve starts to differ from zero around
$\Gamma \approx 1000$. For larger values of $\Gamma$, both shells
do not rotate relative to each other (i.e. they are locked), which
is a typical behavior for a magic number configuration. One can
see that adding small particles influences drastically the
relative angular melting temperature: the value of the coupling
constant $\Gamma$ at which the angular order between both rings is
lost moves to smaller values. This is shown more clearly in
Fig.~\ref{16}: the black squares show the $\Gamma$ value at which
the relative angular diffusion coefficient exceeds the value $100$
as a function of the number of small particles in a cluster
consisting of 16 big particles. One can see that adding even a few
small particles can reduce the critical $\Gamma$ value with a
factor of ten. This leads to a first conclusion that \emph{adding
small particles stabilizes angular order} (i.e. it increases the
rigidity of the cluster). This unexpected increase in angular
order is induced by the vibrations of the small particles. The
vibrating small particles, which are mostly situated between the
inner and outer shell, lock both shells with respect to each other
and stabilize the angular order. Notice the occurrence of a
relatively large critical value of $\Gamma$ for $6$ small
particles. This can be explained in terms of vacancies. Between
the big particles there are only $5$ vacancies. When $6$ small
particles are added to the system, two small particles have to
occupy the same vacancy (see inset of Fig.~\ref{res}(g)). This
distorts the triangular structure and reduces the angular order,
leading to a smaller angular melting temperature. If one increases
the number of small particles further, the angular stabilization
is restored. One can conclude that on average small particles
stabilize the angular order, but the increase in angular melting
temperature with respect to a cluster without small particles
depends on the positioning of the small particles in the
vacancies.

Adding small particles does not only influence the angular melting
temperature, but also the height of the plateau in the angular
diffusion coefficient (shown by the red triangles in Fig.~\ref{16}
at $\Gamma$=20). The height of the plateau indicates how fast the
shells are rotating with respect to each other. There exists an
optimum number of small particles for this angular stabilization
(corresponding to a minimum in this curve), which in this case is
obtained for $N_S=8$ at which the relative angular diffusion
coefficient is reduced with a factor of two with respect to a
system without small particles. Note also that the small dip in
the relative angular diffusion coefficient for the system without
small particles around $\Gamma\approx 8$ is a sign of the
reentrant behavior as studied before in Ref.~\cite{schweigert} and observed in the experiment of R. Bubeck \textit{et al.} \cite{Bubeck}.

Next, we examine how these angular melting properties are modified
for the non-magic cluster with 17 big particles (see right column
in Figs.~\ref{res} and~\ref{17}). Without small particles, the
relative angular order is lost at much smaller temperatures than
for a magic number configuration. Adding one small particle does
not affect this melting temperature substantially for this
particular system, as it will sit in the center of the cluster.
However, adding more than three small particles has an even
stronger influence on the melting temperature for angular melting
(shown by the black squares in Fig.~\ref{17}) in comparison with
the magic-number configuration. Again we found an optimum number
of small particles for angular stabilization, which in this case
also occurs for 8 small particles (see the minimum in the curve
with red triangles in Fig.~\ref{17}).

\subsection{Radial melting}

\begin{figure}
\centering
\includegraphics[width=250pt]{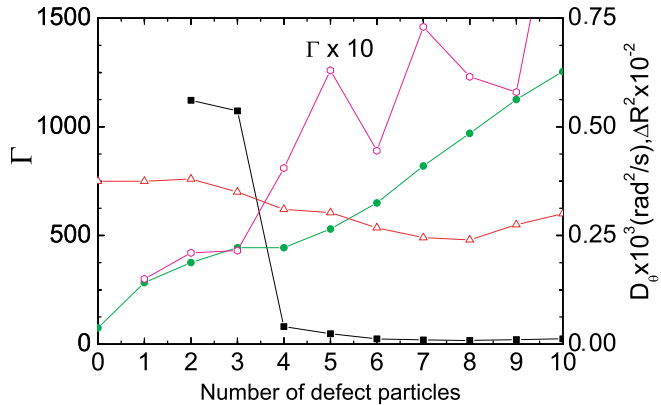} \caption{(color online) The same as Fig.~\ref{16} but now for a cluster consisting of $17$ big particles. } \label{17}
\end{figure}

In order to describe the melting in the radial direction we
calculated $\Delta R^2$ as a function of $\Gamma$. The $\Delta
R^2$ of the big particles in the binary cluster is shown by the
black squares in Fig.~\ref{res}. Notice that the radial melting of
the big particles sets in around $\Gamma=10$ which is independent
of the number of small particles. In the limit of $\Gamma$ to
zero, $\Delta R^2$ approaches $N_B/72$, which is exactly the
theoretical limit for the $\Delta R^2$ of a system of completely
uncorrelated particles in a cavity with hard walls. Notice that
the $\Delta R^2$ curve of the big particles in Fig.~\ref{res},
exibits an increase of the inclination around $\Gamma=10 - 20$ as
a function of the number of small particles. To analyze this
further, we show in Figs.~\ref{16} and~\ref{17} the $\Delta R^2$
of the big particles at $\Gamma=20$ as a function of the number of
small particles (green dots) for $16$ and $17$ big particles,
respectively. This $\Delta R^2$ curve shows a linear increase
which can be understood as follows: as the angular motion of the
big particles is tempered by the small particles, most of the
kinetic energy of the particles is directed into the radial
direction. Summarizing we can conclude that the radial melting
temperature of the big particles is independent of the number of
small particles, but that the thermal fluctuations of the angular
motion of the big particles is compensated by an increase of
$\Delta R^2$.

Comparing the $\Delta R^2$ of the big particles with the $\Delta
R^2$ of the small particles (the red dotted curves in
Fig.~\ref{res}) it is seen that the radial melting of the small
particles sets in at a smaller temperature than the radial melting
of the big particles. This confirms the experimental
observation~\cite{mangold} that the small particles become
delocalized at a larger value of the coupling constant than the
big particles. From the $\Delta R^2$ of the small particles we
notice that the radial melting of the small particles depends on
the number of small particles. Figs.~\ref{16} and~\ref{17} (pink
pentagons) show the critical $\Gamma$ value where $\Delta R^2$
becomes larger than $0.01$ as a function of the number of small
particles for a 16 and 17 particle system, respectively. The
$\Delta R^2$ curves show that the radial melting temperature
increases as function of the number of small particles. This is a
consequence of the induced distortion of the triangular lattice by
the small particles which is proportional to the number of small
particles. The influence of this distortion is clearly visible in
the $\Delta R^2$ curve (pink pentagons in Fig.~\ref{16}), just
like for the angular melting temperature (black squares). For $6$
small particles, at least $2$ particles have to occupy a single
vacancy which distorts significantly the triangular lattice of the
big particles in comparison to the occupancy of only $1$ small
particle per vacancy. This leads also to a weaker pinning of the
small particles and consequently a smaller melting temperature
(i.e. a larger critical coupling constant). Since the interaction
energy of the small particles (with charge = $q_0/8$) is less than
the interaction energy of the big particles, we can expect that
the critical coupling constant at which the small particles melt
is between 8 times (for 1 small particle in the cluster) and 64
times (in the limit that the interaction is completely dominated
by small particles) the critical coupling constant for the big
particles. One can verify in Fig.~\ref{res} that this is indeed
the case.

An unusual behavior is found for the radial melting behavior of
the small particles which is found not to occur in a single step.
The $\Delta R^2$ (red dotted curves in Fig.~\ref{res} for $16$ and
$17$ big particles) increases suddenly at a specific $\Gamma$
value. However, it does not reach its maximum value immediately.
This means that the small particles at this temperature do not
move freely throughout the system, but hop between the vacancies.
By further decreasing $\Gamma$ we even observe a decrease in the
$\Delta R^2$ before it reaches the theoretical limit of $N_B/72$,
in which case the small particles move uncorrelated through the
system. In contrast to the mono dispersive cluster where only an
angular reentrant behavior was observed~\cite{schweigert}, we find
here a \emph{new reentrant behavior} but now in the radial melting
of the small particles. This reentrant melting occurs exactly at
the $\Gamma$ value where the relative angular diffusion
coefficient increases strongly. When at large $\Gamma$ the small
particles temper the relative angular motion of the shells, we
notice now that for smaller $\Gamma$ values it is the complete
angular melting which restricts on its turn the radial motion of
the small particles.

\section{Conclusion}

In conclusion, we investigated the melting behavior of a classical
two-dimensional binary cluster. We showed that defect particles in
such a binary cluster stabilize the angular order of the cluster.
An optimum value for the number of small particles was found for
this increased angular stabilization. This tempering of the
angular motion of the big particles is compensated by an increase
of the radial motion of the big particles. The melting process in
a binary cluster takes place in several steps where first the
small particles and then the big particles become delocalized with
increasing temperature. Due to the radial diffusion of the small
particles, the relative intershell rotation of the big particles
is reduced with respect to a system without small particles.
Further, with an increase of temperature, the diffusion of the big
particles switches on that leads to the stabilization of the
radial motion of the small particles and a reentrant behavior of
the small particles occurs.

\vspace*{1cm}

{\textbf{Acknowledgments}} This work was supported by the Flemish
Science Foundation (FWO-VI). One of us (I.S.) was supported by a
NATO-fellowship.

\end{document}